\documentstyle[preprint,tighten,aps]{revtex}       

\begin{document}

\draft
\title{Solar reaction rates, non-extensivity and quantum uncertainty}
\author{A. Lavagno  and P. Quarati} 
\address{Dipartimento di Fisica, Politecnico di Torino, C.so Duca degli 
Abruzzi 24, I-10129 Torino, Italy}
\address{Istituto Nazionale di Fisica Nucleare, 
Sezioni di Torino e di Cagliari}

\maketitle

\begin{abstract}
We show that in weakly non-ideal plasmas, 
like the solar interior, both 
non-exten\-si\-vi\-ty and quantum uncertainty 
(\`a la Galitskii and Yakimets) 
should be taken into account to derive equilibrium ion 
distribution functions and to estimate nuclear reaction rates  
and solar neutrino fluxes. 

\end{abstract}

\vspace{1cm}

\vspace{1cm}



It is well known that Boltzmann-Gibbs statistics does not describe 
correctly the statistical behavior of particles 
with long-range interaction, non-Markovian memory and  
multifractal boundary conditions \cite{tsa,lang}. 
The treatment in the frame of the extensive description is based on 
short-range interaction, on neglecting surface effects and on fluctuations 
of thermodynamic quantities relatively too small to be observed. 

The solar core plasma represents a system of particles 
that, for many rea\-sons, 
has an equilibrium distribution deviating very slightly
from the Maxwell-Boltzmann (MB) distribution.  
In fact, 
the mean Coulomb energy is not much smaller than the thermal kinetic energy; 
the Debye length is of the order of the inter-ionic distance $a_i$;  
it is not 
possible to clearly separate collective and individual degrees of freedom;   
the presence of the scales of energies of the same size produces deviations 
from the standard statistics \cite{kplb}.

In the region of plasma parameter $\Gamma=Z^2 e^2/(a_i kT)\approx 0.1$,  
to which the solar core plasma belongs, 
the Debye character 
of the screening is maintained but strong interactions at distances of 
the order of 
$n^{-1/3}$ ($n$ is the particle density) 
become the main interactions while involving multiple particles. 
The reaction time to build up screening after a hard collision is the 
inverse of the plasma frequency; 
the collision time is comparable to it. Therefore 
many collisions are necessary before particles loose memory of the initial 
state and the scattering process can not be considered Markovian, 
non-extensive effects should be included \cite{cox}. 
Tsallis non-extensive thermostatistics 
is actually a well-known tool successfully applied to many different 
physical problems and can account quite naturally of 
all the above mentioned features of the solar core plasma 
by means of the entropic parameter $q$ \cite{tsa,kplb}. 

Recently, in Ref.s \cite{sav,sta} 
the Green's function technique has been used 
to exa\-mi\-ne tunneling reaction probability 
and the r\^ole of the many-body effects 
(or quantum uncertainty effect, 
following Galitskii and Yakimets \cite{gal}) 
on the rates of nuclear reactions 
in the laboratory fusion plasma and in the weakly non-ideal sun core.  
The implications of such effects in the evaluation of the solar neutrino 
fluxes are discussed in Ref. \cite{sta} (for an updated reading 
on the solar neutrino problem 
one can refer to Ref.s \cite{caste,basu,dar} 
and to the recent Ref. \cite{giapu} where neutrino decay 
solution is reexamined in the light of the SuperKamiokande data). 

The aim of this paper is twofold:
1) we find a microscopic interpretation of the non-extensive parameter $q$ 
in the solar core  
(in other words, the entity of deviation of the Maxwellian distribution 
of the ionic equilibrium distribution we use in this work) 
in terms of the plasma parameter and the ion correlation parameter; 
2) we study both the non-extensive ($q$) and 
the quantum uncertainty ($Q$) effects \cite{sta,gal}, 
in the calculation of the solar reaction rates and in the solar 
neutrino fluxes.  
The two effects should be simultaneously active if the 
collision frequency is comparable with the plasma frequency; 
this requirement is fulfilled in the solar core.

{\it Electrical microfields and validity of the Tsallis distribution 
in the solar core - }
Basic quantities influencing the microscopic processes 
and the nuclear rates are 
the electrical microfields acting on the nuclei in the plasma: 
the time-spatial fluctuations in the particles positions produce 
specific fluctuations of the microscopic electric field in 
a given point of the plasma. The rates are changed respect to 
the standard ones because of the presence of microfield distributions 
(with energy density 
of the order of $10^{-16}$ MeV/fm$^3$) which 
modifies, under particular conditions, 
the ion distribution tail. 
The influence of microfields on the rates of nuclear tunneling reactions 
has been widely studied since the early work of Holtsmark \cite{hol} 
and later, for instance, in the Ref.s \cite{bara,igle,roma,val}.

In the solar core, the effects of random electric microfields are of 
crucial importance. These microfields   
have in general long-time and long-range 
correlations, can generate anomalous diffusion 
and may be decomposed in three main components: 
1) A slow varying component due to plasma oscillations. 
The particles see this component as an almost constant external mean 
field over several collisions. 
2) A fast random component related to the diffusive cross section 
($\sigma_d \approx 1/v$); this component does not affect 
the distribution that remains Maxwellian.
3) The third component is related to a short-range two-body strong 
Coulomb effective interaction. As we will see later, this is the component 
that alters the distribution.  

Let us derive the ion distribution function to be used  
to calculate tunneling reactions rates in solar plasma, when 
an electric microfield distribution is present. 
The equilibrium distribution we are deriving differs 
from the MB distribution,  
if particular conditions are fulfilled. 
In fact, the presence of the electric microfield average energy 
density, $\langle E^2\rangle$, modifies the stationary solution 
of the Fokker-Planck equation and the ion equilibrium distribution can be 
written as \cite{gola}
\begin{equation}
f(v)=C \exp \left \{ -\int_0^v \frac{m v dv }
{kT + \frac{2}{3} \frac{e^2 \langle E^2 \rangle}{x m \nu^2}}
\right \} \; ,
\label{distri}
\end{equation}
where $\nu$ is the total collision frequency, 
$x$ the elastic energy-transfer coefficient between two particles of the 
plasma, $x=2 m_1 m_2/(m_1+m_2)^2$, 
$m$ the reduced mass, $T$ the temperature, 
$C$ the normalization constant.  \\ 
Defining a critical field 
$E_c=\nu \sqrt{3 x m kT/2 e^2}$,  
we can see from Eq.(\ref{distri}) that 
if $E \ll  E_c$ the distribution 
is Maxwellian whatever be the value of the frequency $\nu$; 
if $E \gg E_c$ and $\nu$ is not a constant but depends on $v$, 
the distribution is a Druyvesteyn like distribution \cite{gola}. 
In the solar core being $E$ not too larger than $E_c$, 
the distribution of Eq.(\ref{distri}) differs slightly from the Maxwellian 
but such small deviation is quite important 
in the evaluation of the nuclear rates.  \\
The condition that $\nu$ be a function of velocity $v$ 
is verified by the fact that the elastic collision cross section is 
$\sigma=\sigma_d+\sigma_0$, where 
$\sigma_d \propto 1/v$ is the elastic diffusion cross section  
and $\sigma_0$ is the enforced Coulomb cross section. 
The total frequency $\nu=\langle \sigma v n\rangle$ satisfies the 
relation $\nu^2=\nu_d^2+\nu_0^2$ without interference 
terms. \\
The expression of $\sigma_0$ is due to 
Ichimaru \cite{ichi} which developed a strict enforcement 
of the Wigner-Seitz ion 
sphere model yielding the elastic cross section $\sigma_0=2\pi(\alpha a)^2$  
where $a$ is the inter-particle distance, 
$\alpha$ is a one dimensional parameter related to the probability that the 
nearest neighbor be at a distance R and therefore related to the 
pair-correlation function $g(R,t)$. 

The explicit expression of the equilibrium 
distribution (\ref{distri}) for the solar interior can be written 
as a function of the kinetic energy $\epsilon_p$ 
\begin{equation}
f(\epsilon_p)=N \exp{\left[ - \varphi \frac{\epsilon_p}{kT}
              - \delta  
\left(\frac{\epsilon_p}{kT}\right)^2 
               \right]}  \, ,
\label{diclay2}
\end{equation}
where $N$ is the normalization constant and 
\begin{equation}
\varphi=\frac{\hat{\varphi}}{1+\hat{\varphi}}\, , \ \ \ \ \ 
\hat{\varphi}=\frac{9}{2} \, x \, 
\frac{n^2 (kT)^2}{Z^2 e^2 \langle E^2\rangle} 
\langle\sigma_d^2\rangle \, ,  
\label{phi}
\end{equation}
\begin{equation}
\delta=\left (\frac{3\langle\sigma_d^2\rangle}{\sigma_0^2}+
\frac{1}{\hat{\delta}} \right)^{-1} \; , \ \ \ \ \ 
\hat{\delta}=\hat{\varphi}  
\frac{\sigma_0^2}{3 \langle\sigma_d^2\rangle}\, . 
\label{delta2}
\end{equation}
In solar interior $\varphi\approx 1$ ($\hat{\varphi}\gg 1$), 
therefore the equilibrium distribution function   
containing random microfields 
with collision frequency depending on the velo\-ci\-ty, 
is given by the Maxwellian distribution times the factor 
$\exp[-\delta (\epsilon_p/kT)^2]$.  
It represents, if we recall $\delta$ as $\delta=(1-q)/2$, 
the approximation of the Tsallis distribution when $q\approx 1$ \cite{tsa}. 
It is remarkable that such distribution has been 
postulated ad hoc by Clayton more than twenty years 
ago in the solution of the solar neutrino problem \cite{clay}. 

By means of Eq.s(\ref{diclay2})-(\ref{delta2}) 
we have established a strict connection between the entity of the 
deviation from the MB in the solar interior, the $\delta$ ($q$) parameter, 
and the elastic diffusion and Coulomb cross sections. 
Expliciting such cross sections in terms of the Ichimaru parameter 
$\alpha$ and the plasma parameter $\Gamma$, 
after straightforward calculations, we obtain that, 
in the small correction limit relevant to the solar core 
($\varphi \approx 1$), the $\delta$ 
parameter can be written as  
\begin{equation}
\label{deltam}
\vert \delta \vert \approx 
\frac{\sigma_0^2}{3 \langle\sigma_d^2\rangle}=12 \, \alpha^4 
\, \Gamma^2 \ll 1 \, .
\label{delgam}
\end{equation}
Eq.(\ref{delgam}) is a crucial result of this paper 
because it establishes that the presence of electric 
microfields implies a deviation from the MB distribution, the 
entity depending on the value of the plasma parameter and the collision 
frequency by means of the $\alpha$ parameter. As we will see, 
very small deviations from the Maxwellian tail can produce strong 
deviations of the reaction rates from their standard values. 

Let us remark that we 
do not discuss the electron screening effect \cite{bahscr}. 
We can realize that at the temperature, elemental densities and plasma 
parameter of the sun core, the rates of the reactions 
could be modified of only few per cent and even less 
by the screening factor $f_0$. 
Its value can be taken equal to unity because its effect on the rates 
is negligible if compared to the depletions (or enhancements) of the rates 
due to other effects (electromagnetic fluctuations in a plasma)  
responsible of the equilibrium distribution function 
we are considering and evaluating in this work.


{\it The quantum uncertainty effect - } 
Galitskii and Yakimets \cite{gal} 
showed that in dense or low temperature plasma, due to quantum 
uncertainty effect, the particle distribution function over momenta 
acquires a power-like tail even under conditions of thermodynamic 
equilibrium  
\begin{equation}
f({\bf p})=\int_{-\infty}^{+\infty} \!\!\!
f(\epsilon,{\bf p}) d\epsilon=f_M({\bf p})+
\frac{\hbar\nu kT}{2\pi\epsilon_p^2} e^{\mu/kT} \; ,
\label{distrim}
\end{equation}
where $f_M({\bf p})$ is the equilibrium MB distribution 
over momenta. \\
The distribution function of Eq.(\ref{distrim}) is an 
approximation of the correct solution, not valid for 
$\hbar\nu/kT\gg 1$ (non-ideal plasma of high density and low temperature), 
and is correct in the case of solar core $\hbar\nu/kT\simeq 0.1$ 
(we remember, as already mentioned, that $\nu\approx \omega_{pl}$). 
At sufficiently high values of ${\bf p}$, 
the magnitude of the tail can be many times greater 
than the Maxwellian distribution. 
Since is the momentum rather than the energy that enters into the 
scattering amplitude in the gas approximation, fusion 
reaction rates in equilibrium sy\-stems can be calculated by averaging the 
reaction cross section $\sigma(\epsilon_p)$ over the momentum 
distribution function, rather than the energy distribution \cite{sav}. 

Let us briefly summarize the results obtained in Ref. \cite{sta} 
taking into account the uncertainty quantum effects $Q$ alone. 
We indicate the corrections derived respect to the standard results 
based on Maxwellian distributions. \\
Although the $pp$ reaction rate should be corrected by a factor 
$r_{11}\simeq 3.5 \cdot 10^{-3}$, (at $r=0.0759 \, R_\odot$)
nevertheless the solar luminosity is maintained unchanged 
respect to the observed value because: 
a) the $^3$He-$^3$He rate should be increased by a factor 
$r_{33}\simeq 4.5 \cdot 10^8$, 
b) the $^3$He-$^4$He rate should be increased by a factor 
$r_{34}\simeq 3.1 \cdot 10^9$, at $r=0.00759 \, R_\odot$ and 
$r_{34}\simeq 5.9 \cdot 10^8$, at $r\simeq 0$, 
c) the $^4$He density is unchanged, 
d) $^3$He density is decreased by $10^8$, 
e) electronic capture (that is not a tunneling reaction) rate by $^7Be$ 
is unchanged, 
f) p-$^7$Be rate is increased by a factor $r_{17}\simeq 4 \cdot 10^5$, 
at $r\simeq 0$, 
g) the density of $^8B$ is increased by $8 \cdot 10^3$, 
h) other reaction rates to burn $^8$B are increased by $\simeq 10^{30}$, 
i) the flux of $^7$Be neutrinos is decreased by a factor $\simeq 50$.

The authors argue that because the mean average molecular weight A is left 
unchanged then, the sound speed does not change respect to the standard 
prediction. This should be proved, because locally in interior shells, 
where several reactions are mostly active, different density (from 
standard) might induce quite different speed of sound and helioseismology 
constraints can be not fulfilled. 
The aim of Ref. \cite{sta} was to check if the calculated 
quantum uncertainty $Q$ 
rates contradict solar data and found it do not. 
Solar interior conditions require the inclusion of the $Q$ corrections. 
Nevertheless, because results are quite different from 
solar standard model (SSM) results 
we will see that only within non-extensive  thermostatistics these 
corrections can be adequately arranged to agree with the experimental results.


{\it Nuclear reaction rates with non-extensive and quantum 
uncertainty corrections - } 
We know that a solution which coincides with a non-extensive 
distribution can be obtained only if the electric 
microfield $E$ is greater than a certain critical value $E_c$, 
and if the frequency of collision depends on $v$.    
This condition is 
verified if the Coulomb collisional cross section,  
which describes the strong part of the collision,  
is a constant: $\sigma_0=2\pi(\alpha a)^2$ 
and is due to the random contribution of the electric microfield. 
The above assumptions do not modify the treatment of Galitskii and Yakimets  
(they consider the Coulomb interaction, but the random part of it is not 
taken into account), rather are responsible of some changements in the 
final rates. 

The non-extensive distribution function, given by Eq.(\ref{diclay2}),  
modifies the reaction rate $R_q$ and can be written as 
\begin{equation}
R_q=R_M \, e^{-\delta \gamma_{ij}} \; , 
\end{equation}
where $R_M$ are the Maxwellian rates, $\delta=(1-q)/2$, 
$\gamma_{ij}=(\epsilon^{ij}_0/kT)^2$, 
$\epsilon^{ij}_0=5.64 (Z_i Z_j \, A_i A_j/(A_i+A_j)\, 
T_c/T )^{1/3}$, 
is the most effective energy and $T_c=1.36$ keV is the temperature at 
the center of the Sun.

If we consider only the quantum effect $Q$,  
the correction $r_{ij}$ to the Maxwellian rate ($R_Q=R_M r_{ij}$) 
can be written as \cite{sav}
\begin{eqnarray}
&&r_{ij}=\frac{3^{19/2}}{8 \pi^{3/2}}\sum_b \frac{h\nu}{kT}
\left (\frac{m_{\rm coll}}{m_r}\right)^{7/2} \frac{e^{\tau_{ij}}}
{\tau_{ij}^8} \; ,\\
&&m_{coll}=\frac{m_r m_b}{m_r+m_b} \; , \ \ \ \ 
m_r=\frac{m_i m_j}{m_i+m_j} \; , \\
&&\tau_{ij}=3 \, (\frac{\pi}{2})^{2/3} 
\left (100 Z_i^2 Z_j^2 \frac{A_i A_j}{A_i+A_j}
T^{-1}_{\rm keV}\right )^{1/3}\; ,
\end{eqnarray}
where $m_b$ is the mass of the background particles colliding 
with tunneling particles.

Including the Tsallis non-extensive effects, 
the total rate (i.e. including both the $q$ and $Q$ effects) is
\begin{equation}
R=R_q+R_Q e^{-\delta \gamma_{ij}^{*}} = 
R_M (e^{-\delta \gamma_{ij}}+
r_{ij} e^{-\delta \gamma_{ij}^{*}}) \; ,
\label{rates}
\end{equation}
where $\gamma^{*}=(\epsilon_Q/kT)^2$ 
and $\epsilon_Q$ is the effective energy of the 
quantum correction. Considering the appropriate expression 
of the collision frequency $\nu$ in terms of $\sigma_0$ and $\sigma_d$, 
we can verify that approximately holds the numerical relation: 
$\epsilon_Q\approx 3 \epsilon_0$. 
Although the values $r_{ij}$ calculated in Ref.\cite{sta} are of order 
$10^8 \div 10^9$, the factor $e^{-\delta \gamma^{*}}$ 
has a large exponent and can strongly suppress the enhancement given by 
$r_{ij}$. 
In fact, the quantum uncertainty 
corrections calculated on the basis of Tsallis distribution 
$r_{ij} e^{-\delta \gamma^{*}}$ are of the same order of 
$e^{-\delta \gamma}$ and even less 
(if $\delta\approx 10^{-2}\div 10^{-3}$, as we can deduce from 
Eq.(\ref{delgam})). 
Solar conditions admit small non-extensive statistics effects 
in the equilibrium distribution, showing very small deviations in the 
Maxwellian tail. 
By using a numerical code based 
on a complete evolutionary stellar model, it has been verified  
that the consequence of this statement are compatible to the 
experimental results on neutrino fluxes \cite{fiore}. 

Let us now define the following quantities:
\begin{eqnarray}
A=\frac{\Phi(Be^7)/\Phi^M(Be^7)}{\Phi(B)/\Phi^M(B)} \, , \ \ 
B=\frac{\Phi(B)}{\Phi^M(B)} \, , 
\ \  C=\frac{\Phi(Be^7)}{\Phi^M(Be^7)} \, , \nonumber 
\end{eqnarray}
and $k_{e7}=z k^M_{17}$, $k_{e,7}=y \, k_{17}$, 
where $k_{ij}$ and $k_{ij}^M$ are respectively the modified and Maxwellian 
rates; the value of the constant $z=227$ can be found, for instance, 
in Ref.\cite{adel}. \\
We can take the usual time-dependent equations of the rates 
with the solar luminosity constraint  
(as reported for instance in Ref.\cite{sta}) 
to derive the steady state solutions 
for the elemental densities. 
Using the expressions of the rates of Eq.(\ref{rates}) with 
quantum uncertainty and non-extensive corrections, 
the following set of equations can be derived:
\begin{eqnarray}
&&A=\frac{C}{B}=\frac{e^{\delta_{17}\gamma^{*}_{17}}}{r_{17}} 
e^{-\delta_{17}\gamma_{17}}\ll r_{17} e^{-\delta_{17}\gamma^{*}_{17}} 
\; , \ \ \
\frac{y}{z}\frac{1}{A}=1 \; ,\nonumber \\
&&\frac{n_{Be^7}}{n^M_{Be^7}}=\frac{n_3}{n_3^M} 
(e^{-\delta_{34}\gamma_{34}}+ r_{34} e^{-\delta_{34}\gamma^{*}_{34}}) 
\, \frac{1}{[1+(e^{-\delta_{17}\gamma_{17}}+ 
e^{-\delta_{17}\gamma^{*}_{17}})/(2z)]} \; . \nonumber 
\end{eqnarray}
A reasonable evaluation of $\alpha$ gives: $\alpha=0.55$, with 
$\Gamma \sim 0.1$  we obtain $q=0.990$ 
($\delta=0.005$) for all components (see the Conclusions for a comment 
on the uncertainty of these values).

If we assume $n_3/n_3^M\simeq 3 \cdot 10^{-3}$  
(let us remark that density is not changed alone, but all the tunneling 
rates are changed consistently solving the set of the mentioned equations)  
we obtain 
\begin{eqnarray}
&&\frac{\Phi (Be^7)}{\Phi^M (Be^7)}=\frac{1}{50} \ \ \ , \ \ \ 
\frac{\Phi (B)}{\Phi^M (B)}\approx \frac{1}{2} \; ,\nonumber \\
&&{\rm Gallium}=81 \; {\rm SNU}\; , \ \ \  
{\rm Chlorine}=2.8\; {\rm SNU} \; ,\nonumber
\end{eqnarray}
$\Phi(pp)$ and luminosity are practically unchanged respect 
to the SSM values. 

The CNO rates that are strongly enhanced by the quantum uncertainty 
effect are remarkably 
reduced by the factor $e^{-\delta \gamma^{*}}$.
We have introduced the assumption that $n_3$ is of the order of about 
$10^{-3}$ reduced 
respect SSM and consistent with the set of time-dependent equations of the 
rates. 
This value is wi\-thin the constraints actually imposed by 
helioseismology because in the region $r/R_{sun}< 0.2$ the value of $n_3$ 
can be submitted to a large variability \cite{brun}. 
The results presented, that are in very good agreement 
with the experimental data \cite{caste,basu,dar}, do not pretend 
to demonstrate that the solar neutrino problem is solved, but rather that 
a path to solution along this procedure can be found.


{\it Conclusions - } 
In a weakly non-ideal plasma, like the solar interior, ions and electrons 
behave like quasi-particles and, due to the quantum uncertainty, 
momentum and energy of colliding particles are independent variables. 
This many-body effect is responsible of an 
additional power-law tail to the Maxwellian momentum equilibrium 
distribution, as discussed by Galitskii and Yakimets \cite{gal}, 
while the energy particle distribution remains Maxwellian. 
The central and more important quantity that enters into the 
quantum correction of the distribution is the 
elastic collision frequency which, 
in these weakly non-ideal plasmas, has a value of the same order as 
the collective plasma frequency.
Being such a value of collision frequency in 
systems of particles with long-range 
interactions related also to memory effects with long-time tails, 
the equilibrium non-extensive 
statistics appears to be the most suitable statistics to use 
and it can take into account, in a very natural way, 
both quantum and non-extensive effects. \\
Savchenko \cite{sav} has shown that the rigorous procedure 
of averaging of tunneling probability, based on Green function technique, 
gives results 
nearly coincident with the results obtained by the simple averaging.
Therefore, we can show that, 
by using the modified distribution function, Eq.(\ref{diclay2}), 
(which corresponds to the equilibrium Tsallis distribution when deviations 
from standard statistics are small) instead of the MB one, 
the strong increase of the nuclear reaction rates 
caused by quantum uncertainty effects alone is greatly (or fully) reduced. 
We can confirm the validity of the Tsallis statistics to describe 
weakly non-ideal plasmas such as the solar core and of its use to 
calculate solar neutrino fluxes towards a closer 
agreement to the experimental measurements than standard calculations. 
In fact, the calculated fluxes are in good agreement with the experimental 
results using the reaction rates of Eq.(\ref{rates}) with quantum uncertainty 
and non-extensive corrections and a $^3He$ density (in the core) 
$10^{-3}$ times the value usually taken in SSM. This figure is 
consistent with helio\-sei\-smo\-logy constraints. 
Finally, we have justified the entropic parameter $q$ ($\delta$) 
of the non-extensive 
statistics because we have related this quantity to the electric microfields 
distribution in the plasma (or explicitly, to the plasma parameter 
$\Gamma$ and to $\alpha$, the ion correlation function).  
Therefore the quantity $q$ cannot be considered a free parameter. However, 
the evaluation of $q$ (or $\delta$) by means of Eq.(\ref{deltam}) 
is not absent of some uncertainty because the quantity $\alpha$ 
($\delta$ depends on the fourth power of $\alpha$) is still to be 
deeply analyzed and evaluated, in spite of the many efforts spent in the 
past on space-time correlation functions of ions in stellar plasma. 
Our results are, of course, not conclusive on the pro\-blem of 
solar neutrinos, rather they represent an indication that, possibly, 
an astrophysical solution can be found once nuclear reactions in weakly 
non-ideal astrophysical plasma be fully understood or measured  
in laboratories with ad hoc experimental facilities. 

This work is supported in part by MURST.

{}

\end{document}